\documentclass[aps,pra,twocolumn,amsmath,amssymb,superscriptaddress,reprint,longbibliography]{revtex4-1}

\usepackage[utf8]{inputenc}
\usepackage{graphicx}
\usepackage[colorlinks=true,citecolor=blue]{hyperref}
\usepackage{units}

\newcommand{\bra}[1]{\langle #1|}
\newcommand{\ket}[1]{|#1\rangle}

\DeclareMathOperator{\sign}{sign}
\newcommand{\up}{\uparrow}
\newcommand{\down}{\downarrow}
\renewcommand{\vec}[1]{\mathbf{#1}}
\newcommand{\kB}{k_\text{B}}
\newcommand{\kBT}{k_\text{B}T}

\begin{document}
\title{Phase-dependent heat and charge transport through superconductor-quantum dot hybrids}
\author{Mathias Kamp}
\affiliation{Theoretische Physik, Universität Duisburg-Essen and CENIDE, D-47048 Duisburg, Germany}
\author{Björn Sothmann}
\affiliation{Theoretische Physik, Universität Duisburg-Essen and CENIDE, D-47048 Duisburg, Germany}
\date{\today}

\begin{abstract}
We analyze heat and charge transport through a single-level quantum dot coupled to two BCS superconductors at different temperatures to first order in the tunnel coupling. In order to describe the system theoretically, we extend a real-time diagrammatic technique that allows us to capture the interplay between superconducting correlations, strong Coulomb interactions and nonequilibrium physics. We find that a thermoelectric effect can arise due to the superconducting proximity effect on the dot. In the nonlinear regime, the thermoelectric current can also flow at the particle-hole symmetric point due to a level renormalization caused by virtual tunneling between the dot and the leads. The heat current through the quantum dot is sensitive to the superconducting phase difference. In the nonlinear regime, the system can act as a thermal diode.
\end{abstract}

\maketitle
\section{\label{sec:intro}Introduction}
Understanding, manipulating and managing heat flows at the nanoscale is of crucial importance for modern electronics where Joule heating constitutes a major nuisance in the operation of computer chips. Heat transport can occur via electrons~\cite{giazotto_opportunities_2006}, phonons~\cite{li_colloquium:_2012} and photons~\cite{meschke_single-mode_2006,ronzani_tunable_2018}.
A promising direction to achieve control over thermal transport by electrons is phase-coherent caloritronics~\cite{martinez-perez_coherent_2014,fornieri_towards_2017} in superconducting circuits. Phase-coherent caloritronics is based on the observation that not only the charge current depends on the phase difference across the junction via the Josephson effect~\cite{josephson_possible_1962} but that also the heat current is sensitive to the phase difference~\cite{maki_entropy_1965,maki_entropy_1966,guttman_phase-dependent_1997,guttman_thermoelectric_1997,guttman_interference_1998,zhao_phase_2003,zhao_heat_2004}. The phase-dependent contribution to the heat current arises from Andreev like processes where an incident electronlike quasiparticle above the superconducting gap is reflected as a holelike quasi particle and vice versa.

Recently, phase-coherent heat transport in superconducting circuits has been observed experimentally~\cite{giazotto_josephson_2012}. The possibility to control heat currents via magnetic fields has led to a number of proposals for phase-coherent caloritronic devices such as heat interferometers~\cite{giazotto_phase-controlled_2012,martinez-perez_fully_2013} and diffractors~\cite{giazotto_coherent_2013,guarcello_coherent_2016}, thermal rectifiers~\cite{giazotto_thermal_2013,martinez-perez_efficient_2013,fornieri_normal_2014,fornieri_electronic_2015}, transistors~\cite{giazotto_proposal_2014,fornieri_negative_2016}, switches~\cite{sothmann_high-efficiency_2017} and circulators~\cite{hwang_phase-coherent_2018}, thermometers~\cite{giazotto_ferromagnetic-insulator-based_2015,guarcello_non-linear_2018} as well as heat engines~\cite{marchegiani_self-oscillating_2016,hofer_autonomous_2016,vischi_coherent_2018} and refrigerators~\cite{solinas_microwave_2016,marchegiani_-chip_2017}.
Experimentally, heat interferometers~\cite{giazotto_josephson_2012,fornieri_nanoscale_2016,fornieri_0_2017}, the quantum diffraction of heat~\cite{martinez-perez_quantum_2014}, thermal diodes~\cite{martinez-perez_rectification_2015} and a thermal router~\cite{timossi_phase-tunable_2018} have been realized so far.
Apart from potential applications in caloritronic and thermal logic~\cite{paolucci_phase-tunable_2018}, phase-coherent heat transport can also serve as a diagnostic tool that allows one, e.g., to probe the existence of topological Andreev bound states~\cite{sothmann_fingerprint_2016}.

\begin{figure}
    \includegraphics[width=\columnwidth]{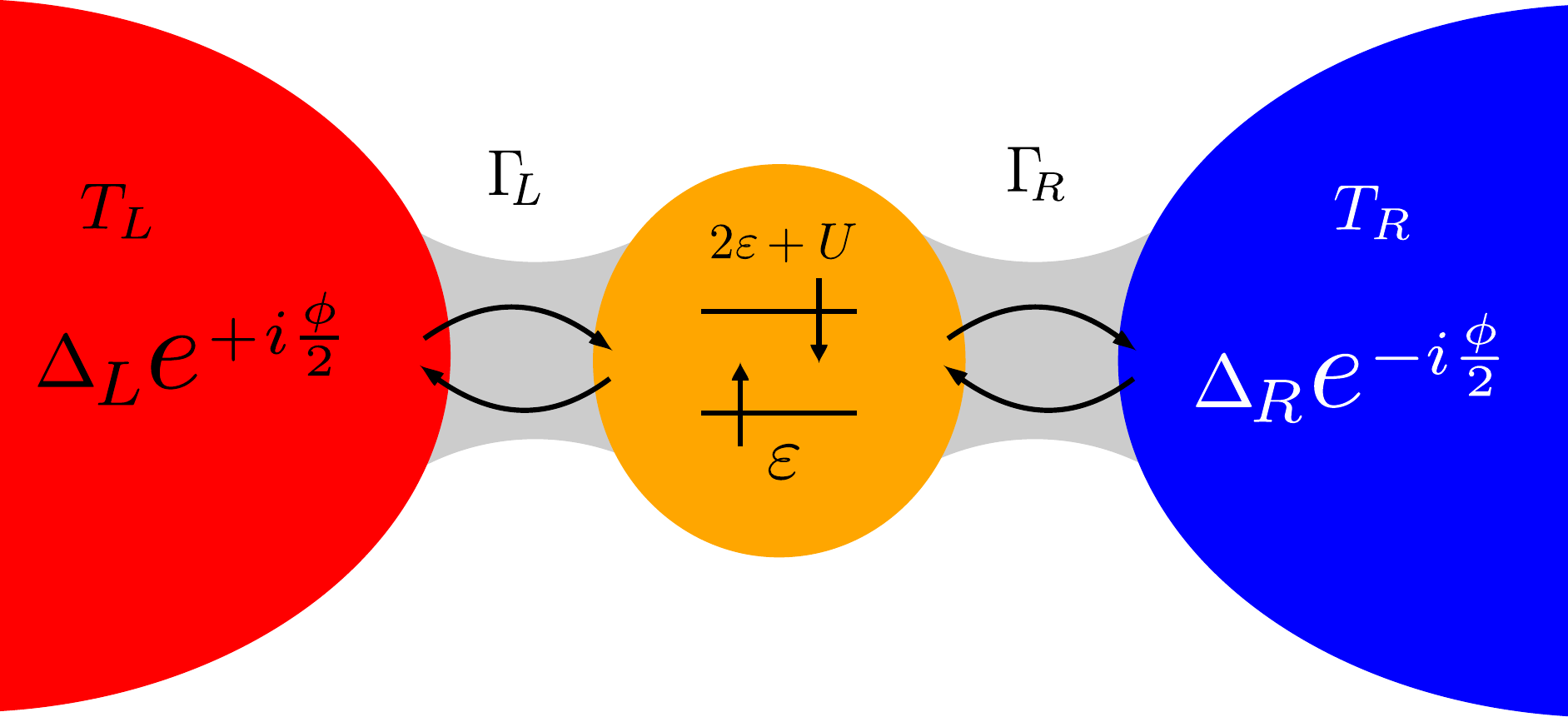}
	\caption{\label{fig:model}Schematic sketch of our setup. A single-level quantum dot is tunnel coupled to two superconducting electrodes at temperatures $T_\text{L}$ and $T_\text{R}$.}
\end{figure}
So far, the theoretical and experimental investigation of phase-coherent heat transport has been restricted to systems such as tunnel barriers and point contacts where the effects of electron-electron interactions can be neglected. While such setups already offer a lot of interesting physics, this raises the question of how Coulomb interactions can affect phase-dependent heat currents. In this paper, we address this important question by analyzing phase-coherent heat and charge transport through a thermally biased hybrid structure consisting of a strongly interacting single-level quantum dot tunnel coupled to superconducting electrodes, cf. Fig.~\ref{fig:model}.

Superconductor-quantum dot hybrids have received a lot of attention, see Ref.~\cite{de_franceschi_hybrid_2010} and~\cite{martin-rodero_josephson_2011} for recent reviews on experiments and theory, respectively. In particular, there are investigations of the Josephson effect through quantum dots~\cite{van_dam_supercurrent_2006,jarillo-herrero_quantum_2006,jorgensen_critical_2007,baba_superconducting_2015,szombati_josephson_2016,probst_signatures_2016}, multiple Andreev reflections~\cite{levy_yeyati_resonant_1997,buitelaar_multiple_2003,cuevas_full_2003,nilsson_supercurrent_2011,rentrop_nonequilibrium_2014,hwang_hybrid_2016}, the interplay between superconducting correlations and the Kondo effect~\cite{clerk_loss_2000,buitelaar_quantum_2002,avishai_superconductor-quantum_2003,eichler_even-odd_2007,lopez_josephson_2007,karrasch_josephson_2008}, the generation of unconventional superconducting correlations in quantum dots~\cite{sothmann_unconventional_2014,kashuba_majorana_2017,weiss_odd-triplet_2017,hwang_odd-frequency_2017}, Cooper pair splitting~\cite{recher_andreev_2001,hofstetter_cooper_2009,herrmann_carbon_2010,hofstetter_finite-bias_2011,das_high-efficiency_2012,schindele_near-unity_2012} and the generation of Majorana fermions~\cite{leijnse_parity_2012,sothmann_fractional_2013,fulga_adaptive_2013,deng_majorana_2016}. 
Thermoelectric effects in superconductor-quantum dot hybrids have been studied in the absence of Coulomb interactions~\cite{kleeorin_large_2016}.
Here, we use a superconductor-quantum dot hybrid as a playground to investigate the interplay between superconductivity, strong Coulomb interactions and thermal nonequilibrium.
Compared to tunnel junctions, quantum dots offer additional tunability of their level position by gate voltages. 
We extend a real-time diagrammatic approach~\cite{konig_zero-bias_1996,konig_resonant_1996,schoeller_transport_1997,konig_quantum_1999,governale_real-time_2008,governale_erratum:_2008} to describe thermally-driven transport which allows us to treat Coulomb interactions exactly and to perform a systematic expansion in the tunnel coupling between the dot and the superconducting leads. It allows for a treatment of superconducting correlations induced on the dot via the proximity effect and captures renormalization effects due to virtual tunneling which affect transport already in lowest order of perturbation theory. 
We evaluate charge and heat currents both in linear and nonlinear response. In particular, we find a thermoelectric effect in the vicinity of the particle-hole symmetric point which arises from the proximity effect. Furthermore, our device can act as an efficient thermal diode in nonlinear response.

The paper is organized as follows. In Sec.~\ref{sec:model}, we introduce the model of our setup. The real-time diagrammatic transport theory used to investigate transport is introduced in Sec.~\ref{sec:method}. We present the results of our analysis in Sec.~\ref{ssec:linear} for the linear and in Sec.~\ref{ssec:nonlinear} for the nonlinear transport regime. Conclusions are drawn in Sec.~\ref{sec:conclusion}.

\section{\label{sec:model}Model}
We consider a single-level quantum dot weakly tunnel coupled to two conventional superconducting electrodes. Both superconductors are kept at the same chemical potential $\mu=0$ but at different temperatures $T_\text{L}$ and $T_\text{R}$ resulting in a nonequilibrium situation.
The system is described by the total Hamiltonian
\begin{equation}
	H=\sum_{\eta=\text{L,R}}\left(H_\eta+H_{\text{tun},\eta}\right)+H_\text{dot},
\end{equation}
where $\eta$ denotes the left (L) and right (R) superconductor. The superconducting leads are characterized by the mean-field BCS Hamiltonian
\begin{equation}\label{eq:BCS}
	H_\eta=\sum_{\vec k\sigma} \varepsilon_{\eta\vec k} a^\dagger_{\eta\vec k\sigma} a_{\eta\vec k\sigma}+\Delta_\eta e^{i\phi_\eta }\sum_{\vec k}a_{\eta -\vec k\up}a_{\eta \vec k\down}+\text{H.c.},
\end{equation}
where $a_{\eta\vec k\sigma}^\dagger$ ($a_{\eta\vec k\sigma}$) denotes the creation (annihilation) operator of an electron with momentum $\vec k$, spin $\sigma$ and kinetic energy $\varepsilon_{\eta \vec k}$ in lead $\eta$. The second term on the right-hand side of Eq.\eqref{eq:BCS} describes the BCS pair interaction on a mean-field level. The two superconducting order parameters are characterized by their absolute value $\Delta_\eta$ and their phase $\phi_\eta$. The temperature dependence of $\Delta_\eta$ is determined by the solution of the self-consistency equation for the order parameter which can be found only numerically. However, it can be approximated with an accuracy of better than 2\% by
\begin{equation}
	\Delta_\eta(T_\eta)=\Delta_{0} \tanh \left(1.74 \sqrt{\frac{T_{c}}{T_\eta}-1}\right),
\end{equation}
in the whole temperature range from 0 to the critical temperature $T_{c}$. The latter is connected to the superconducting order parameter at zero temperature via $\kB T_{c}\approx 0.568 \Delta_{0}$.

The single-level quantum dot is described by the Hamiltonian
\begin{equation}
	H_\text{dot}=\sum_\sigma \varepsilon c_\sigma^\dagger c_\sigma+U c_\up^\dagger c_\up c_\down^\dagger c_\down.
\end{equation}
While the first term describes the energy of the dot level $\varepsilon$ that can be tuned by applying a gate voltage, the second term denotes the Coulomb interaction that has to be supplied in order to occupy the dot with two electrons at the same time. We remark that the dot spectrum is particle-hole symmetric at $\varepsilon=-U/2$. For later convenience, we introduce the detuning $\delta=2\varepsilon+U$ from the particle-hole symmetric point.

The tunneling Hamiltonian which couples the dot to the superconducting leads is given by 
\begin{equation}
	H_\text{tun}=\sum_{\eta \vec k\sigma}t_\eta a_{\eta\vec k\sigma}^\dagger c_\sigma+\text{H.c.}
\end{equation}
Here, $t_\eta$ denotes a tunnel matrix element which we assume to be energy and momentum independent. It is connected to the tunnel coupling strength $\Gamma_\eta=2\pi|t_\eta|^2\rho_\eta$ where $\rho_\eta$ denotes the density of states of lead $\eta$ in the normal state.

\section{\label{sec:method}Real-time diagrammatic transport theory}
In order to describe transport through the quantum-dot setup, we make use of a real-time diagrammatic technique~\cite{konig_zero-bias_1996,konig_resonant_1996,schoeller_transport_1997,konig_quantum_1999} for systems with superconducting leads with a finite gap~\cite{governale_real-time_2008,governale_erratum:_2008}. It allows us to treat nonequilibrium physics, superconducting correlations and strong Coulomb interactions exactly while performing a systematic expansion in the dot-lead couplings. In the following, we are going to extend this diagrammatic framework to allow for the calculation of thermally-driven charge and heat currents through quantum dot-superconductor hybrids on equal footing.

The central idea of the diagrammatic approach is to integrate out the noninteracting leads and to describe the remaining quantum dot system by its reduced density matrix. The reduced density matrix $\rho_\text{red}$ has matrix elements $P^{\chi_1}_{\chi_2}=\bra{\chi_1}\rho_\text{red}\ket{\chi_2}$. For the system under investigation, the nonvanishing density matrix elements are given by the probability to find the quantum dot empty, $P_0$, occupied with a single electron with spin $\sigma$, $P_\sigma$, or doubly occupied, $P_d$. Furthermore, the coupling to the superconductors gives rise to finite off-diagonal density matrix elements $P^d_0$ and $P^0_d$ that describe the coherent superposition of the dot being empty and occupied with two electrons. The generation of these coherent superpositions is a hallmark of the superconducting proximity effect on the quantum dot.

The time evolution of the reduced density matrix is given by the generalized master equation which in the stationary limit reads
\begin{equation}
	0=-i(E_{\chi_1}-E_{\chi_2})P^{\chi_1}_{\chi_2}+\sum_{\chi_1'\chi_2'}W^{\chi_1\chi_1'}_{\chi_2\chi_2'}P^{\chi_1'}_{\chi_2'},
\end{equation}
where $E_\chi$ is the energy of the many-body dot state $\chi$. The first term describes the coherent evolution of the dot states. The second term arises due to the dissipative coupling to the superconductors. The generalized transition rates $W^{\chi_1\chi_1'}_{\chi_2\chi_2'}$ are obtained from irreducible self-energy diagrams of the dot propagator on the Keldysh contour~\cite{governale_real-time_2008,governale_erratum:_2008}, cf. also Appendix~\ref{app:RTD} for a detailed explanation of the connection between diagrams and physical processes. By expanding both the density matrix elements as well as the generalized transition rates up to first order in the tunnel couplings, we find that the coherent superpositions $P^0_d$ and $P^d_0$ are finite to lowest order in $\Gamma_\eta$ only if the empty and doubly occupied dot states are nearly degenerate, $\delta\lesssim\Gamma_\eta$~\cite{sothmann_influence_2010}. For this reason, we are going to restrict ourselves to the analysis of transport in the vicinity of the particle-hole symmetric point to first order in the tunnel coupling in the following.

The generalized master equation can be brought into a physically intuitive form by introducing the probabilities to find the dot occupied with an even and odd number of electrons,
\begin{equation}
	\vec P=\left(\begin{array}{c} P_\text{e}\\P_\text{o} \end{array}\right)=\left(\begin{array}{c} P_0+P_d \\ P_\up+P_\down \end{array}\right),
\end{equation}
as well as a pseudospin degree of freedom that characterizes the coherences between empty and doubly occupied dot and, thus, the superconducting proximity effect on the quantum dot
\begin{align}
	I_x&=\frac{P^0_d+P^d_0}{2},\\
	I_y&=i\frac{P^0_d-P^d_0}{2},\\
	I_z&=\frac{P_0-P_d}{2}.
\end{align}

The generalized master equation can be decomposed into one set of equations that arises from the time evolution of the dot occupations and another set due to the pseudospin. The former is given by
\begin{equation}\label{eq:MEP}
    0
	=\sum_\eta \left[\left(\begin{array}{cc} -Z^-_\eta & Z^+_\eta \\ Z^-_\eta & -Z^+_\eta \end{array}\right)\vec P
	+
	  \left(\begin{array}{c} 4X^-_\eta \\ -4X^-_\eta \end{array}\right)\vec I\cdot \vec n_\eta\right],
\end{equation}
where
\begin{equation}
	X^\pm_\eta=\pm\frac{\Gamma_\eta}{\hbar}\frac{\Delta_\eta \Theta(U/2-\Delta_\eta)}{\sqrt{(U/2)^2-\Delta_\eta^2}}f_\eta(\pm U/2),
\end{equation}
\begin{equation}
	Z^\pm_\eta=\frac{\Gamma_\eta}{\hbar}\frac{U\Theta(U/2-\Delta_\eta)}{\sqrt{(U/2)^2-\Delta_\eta^2}}f_\eta(\pm U/2),
\end{equation}
with the Fermi function $f_\eta(\omega)=[\exp(\omega/(\kBT_\eta))+1]^{-1}$. $\vec n_\eta=(\cos\phi_\eta,\sin\phi_\eta,0)$ denotes a unit vector whose direction is determined by the phase of the superconducting order parameters. Interestingly, in Eq.~\eqref{eq:MEP} the dot occupations are coupled to the pseudospin degree of freedom. This is in direct analogy to the case of a quantum dot weakly coupled to ferromagnetic electrodes where the dot occupations are linked to the spin accumulation in the dot~\cite{konig_interaction-driven_2003,braun_theory_2004}. The second set of equations is given by a Bloch-type equation for the pseudospin,
\begin{equation}
	0
	=\left(\frac{d\vec I}{dt}\right)_\text{acc}-\frac{\vec I}{\tau_\text{rel}}+\vec I\times\vec B.
\end{equation}
The first term,
\begin{equation}
	\left(\frac{d\vec I}{dt}\right)_\text{acc}=\sum_\eta \left(X^-_\eta P_\text{e}+X^+_\eta P_\text{o}\right)\vec n_\eta,
\end{equation}
describes the accumulation of pseudospin on the dot due to tunneling in and out of electrons. The second term characterizes the relaxation of the pseudospin due to electron tunneling on a time scale given by $\tau_\text{rel}^{-1}=\sum_\eta Z^-_\eta$. Finally, the last term gives rise to a precession of the pseudospin in an effective exchange field,
\begin{equation}
	\vec B=B_\text{L}\vec n_\text{L}+B_\text{R}\vec n_\text{R}+\delta \vec e_z,
\end{equation}
which arises from virtual charge fluctuations on the dot as well as from a detuning away from the particle-hole symmetric point.
The exchange field contribution from the two leads is given by
\begin{equation}\label{eq:Bex}
	B_\eta=\frac{2\Gamma_\eta}{\pi\hbar}\int'd\omega\frac{\Delta_\eta\Theta(|\omega|-\Delta_\eta)}{\sqrt{\omega^2-\Delta_\eta^2}}\frac{f_\eta(\omega)}{\omega+U/2}\sign\omega,
\end{equation}
where the prime indicates the principal value. The integral can be solved analytically as an infinite sum over Matsubara frequencies, see Appendix~\ref{app:Bex} for details.
The interplay of pseudospin accumulation, pseudospin relaxation and pseudospin precession in the exchange field leads to a nontrivial pseudospin dynamics on the dot which acts back on the dot occupations via Eq.~\eqref{eq:MEP}. It is this nontrivial pseudospin behavior that gives rise to interesting transport properties of the system under investigation.

The charge on the quantum dot is related to the $z$ component of the pseudospin via $Q_\text{dot}=e(1-2I_z)$. This allows us to connect the time evolution of $I_z$ directly to the charge current flowing between the dot and lead $\eta$ via
\begin{equation}\label{eq:Ic}
	I^e_\eta=-2e(Z_\eta^-I_z-I_xB_{\eta,y}+I_yB_{\eta,x}).
\end{equation}
We remark that the real-time diagrammatic approach conserves charge currents automatically. Therefore, we define $I^e=I^e_\text{L}=-I^e_\text{R}$ in the following. In analogy to the charge, we can relate the average dot energy to the probability to find the dot with an odd occupation, $E_\text{dot}=-UP_\text{o}/2$, to derive for the heat current between the dot and lead $\eta$
\begin{equation}
	I^h_\eta=-\frac{U}{2}\left(Z^+_\eta P_\text{o}-Z^-_\eta P_\text{e}+4X^-_\eta \vec I\cdot\vec n_\eta\right).
\end{equation}
We remark that in the absence of any bias voltage there is no Joule heating and, hence, heat and energy currents are equal to each other. This implies that heat currents are conserved such that we can define $I^h=I^h_\text{L}=-I^h_\text{R}$.

\section{\label{sec:results}Results}
In this section, we are going to analyze the charge and heat currents flowing through the system in response to an applied temperature bias. We will first focus on the linear-response regime and then turn to a discussion of nonlinear transport.

\subsection{\label{ssec:linear}Linear response}
For the sake of concreteness, we consider a symmetric quantum-dot setup. To this end, we define the temperatures of the superconducting leads as $T_\eta=T+\Delta T_\eta$ with the reference temperature $T$ and the temperature bias $\Delta T_\text{L}=-\Delta T_\text{R}\equiv \Delta T/2$. The tunnel couplings are chosen equal, $\Gamma_\text{L}=\Gamma_\text{R} \equiv\Gamma/2$. Furthermore, we assume that the two superconducting order parameters have the same absolute value, $\Delta_\text{L}(T)=\Delta_\text{R}(T)=\Delta$, and set their phases as $\phi_\text{L}=-\phi_\text{R}\equiv\phi/2$.

To zeroth order in $\Delta T$, i.e., in thermal equilibrium the occupation probabilities of the dot are given by Boltzman factors $P_\chi^{(0)}\propto e^{-E_\chi/\kBT}$. At the same time, the pseudospin accumulation on the dot vanishes exactly. In consequence, there is no charge and heat current flowing through the system.
Since we consider only tunnel events that are first order in the tunnel coupling, there is no supercurrent through the quantum dot~\cite{governale_real-time_2008}. The latter would manifest itself as a phase-dependent equilibrium contribution to the charge current. It requires, however, the coherent transfer of Cooper pairs through the dot and, hence, higher order tunnel processes.

\begin{figure}
	\includegraphics[width=\columnwidth]{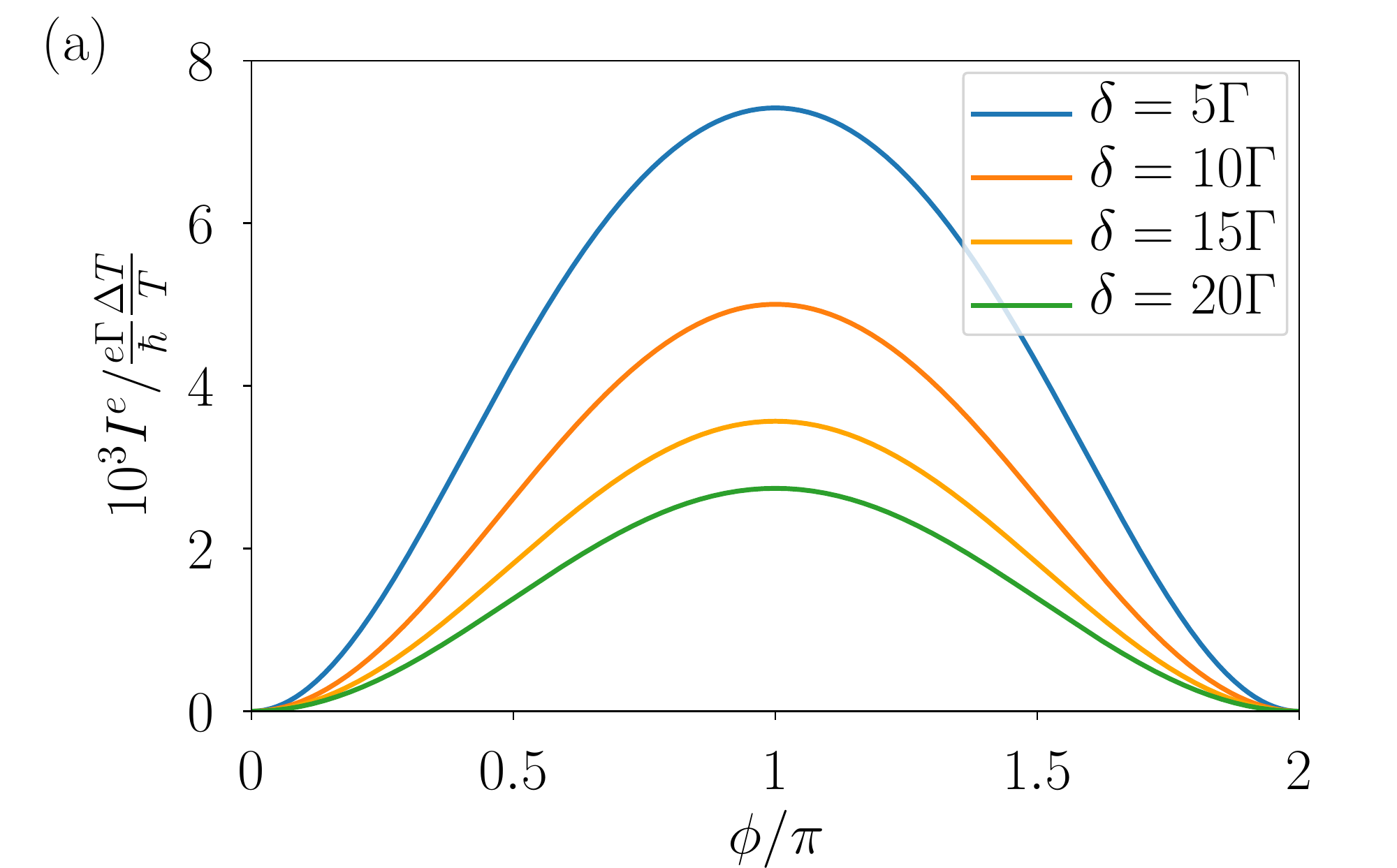}
	\includegraphics[width=\columnwidth]{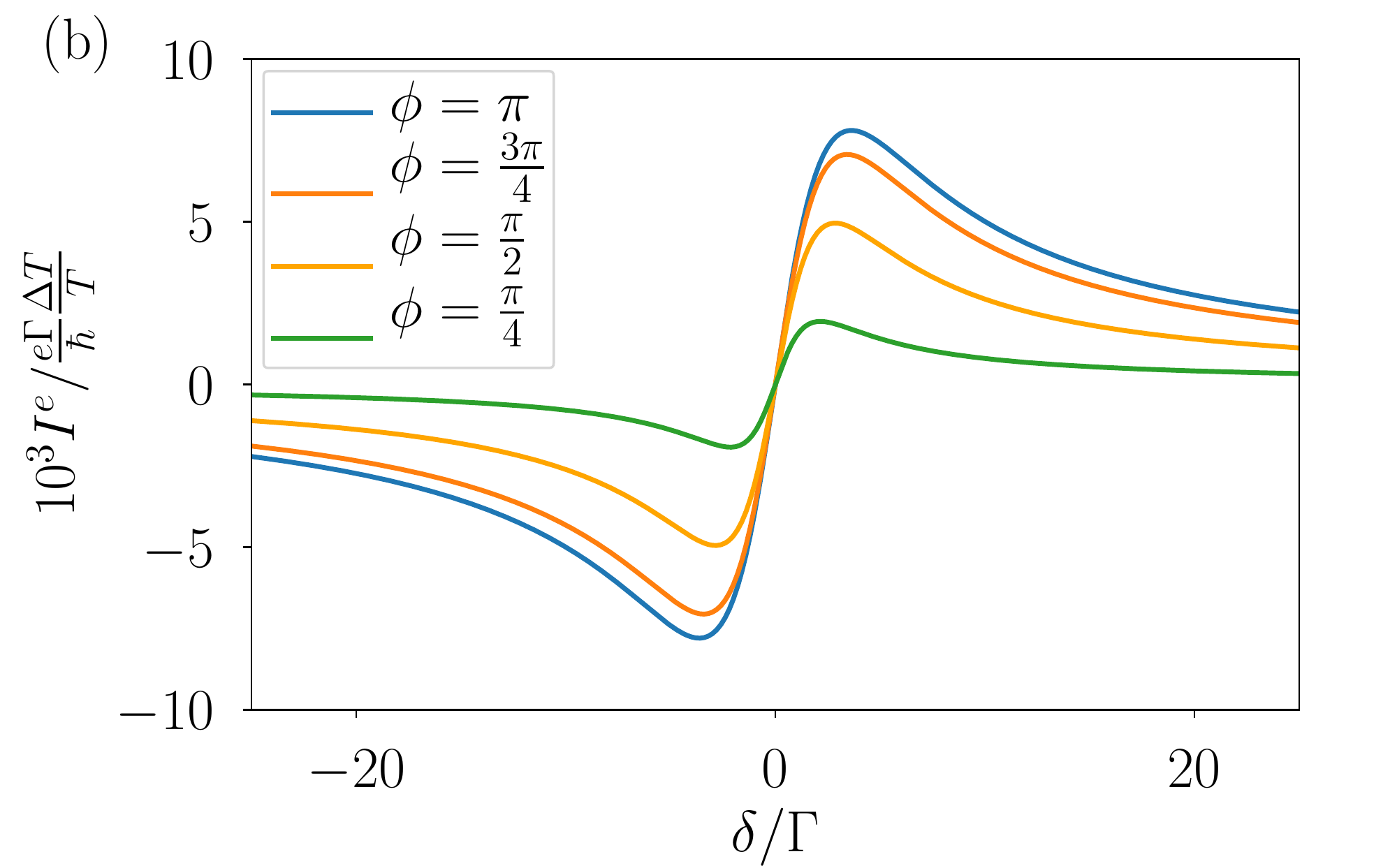}
	\caption{\label{fig:Iclin}Linear-response charge current $I^e$ as a function of (a) phase difference $\phi$ and (b) detuning $\delta$. Parameters are $U=4\kBT$ and $\Delta=1.75\kBT$.}
\end{figure}
A finite temperature bias $\Delta T$ generates a finite pseudospin accumulation on the dot. To first order in $\Delta T$ the accumulation is along the direction $\vec n_\text{L}-\vec n_\text{R}$, i.e., a finite pseudospin component $I^{(1)}_y$ is generated due to nonequilibrium tunneling of electrons. The magnitude of the pseudospin accumulation is limited by the pseudospin relaxation term $-\vec I/\tau_\text{rel}$. In addition, the effective exchange field $\vec B$ gives rise to a precession of the accumulated pseudospin and leads to finite pseudospin components $I^{(1)}_x$ and $I^{(1)}_z$.
According to Eq.~\eqref{eq:Ic}, the pseudospin accumulation leads to a finite charge current given by
\begin{equation}\label{eq:Ielin}
	I^e=-e\frac{2B_0 X_1^-Z_0^-\sin^2\frac{\phi}{2}}{Z_0^-\frac{\delta}{\hbar}+2[(Z_0^-)^2+B_0^2\cos^2\frac{\phi}{2}]\tan\beta}\frac{\Delta T}{T}.
\end{equation}
Here, we introduced the expansions
\begin{align}
	X^\pm_\eta&=X^\pm_0+X^\pm_1\frac{\Delta T_\eta}{T}+\mathcal O(\Delta T_\eta^2),\\
	Z^\pm_\eta&=Z^\pm_0+Z^\pm_1\frac{\Delta T_\eta}{T}+\mathcal O(\Delta T_\eta^2),\\
	B_\eta&=B_0+B_1\frac{\Delta T_\eta}{T}+\mathcal O(\Delta T_\eta^2),
\end{align}
as well as the angle $\beta=\arctan (I^{(1)}_y/I^{(1)}_x)$ which can be written as
\begin{equation}
	\tan\beta=\frac{2\hbar}{\delta Z^-_0}\left[(Z_0^-)^2-4(X_0^-)^2\cos^2\frac{\phi}{2}\right].
\end{equation}
The thermoelectric charge current Eq.~\eqref{eq:Ielin} arises in the vicinity of the particle-hole symmetric point. 
It relies crucially on the superconducting proximity effect and the resulting pseudospin accumulation on the dot because the Fermi functions in the generalized transition rates $W^{\chi_1\chi_1'}_{\chi_2\chi_2'}$ are evaluated at the particle-hole symmetric point $\delta=0$ and, therefore, do not lead to any thermoelectric effect. It is, thus, the pseudospin accumulation that introduces a nontrivial $\delta$ dependence into the master equation via the effective exchange field $\vec B$. In consequence, the thermoelectric current vanishes for $\Delta\to0$, i.e., in the absence of superconductivity in the leads.

In Fig.~\ref{fig:Iclin} (a), the charge current is shown as a function of the phase difference $\phi$. At zero phase difference, the charge current vanishes independently of the detuning $\delta$ because there is no pseudospin accumulation on the quantum dot. In contrast, at $\phi=\pi$ the charge current becomes maximal due to the strong pseudospin accumulation on the dot. 
Figure~\ref{fig:Iclin} (b) shows the charge current as a function of the detuning $\delta$. For $\delta=0$ the charge current vanishes due to particle-hole symmetry. For positive (negative) values of the detuning the charge current takes positive (negative) values indicating electron (hole) transport. The maximal current occurs for a phase difference of $\phi=\pi$ and detuning $\delta=\pm2\hbar Z^-_0$ and takes the value $I^e=-(e B_0 X_1^-\Delta T)/(2Z_0^- T)$. The maximum current is exponentially suppressed in $U/(\kBT)$ due to the requirement of thermally excited quasiparticles. At the same time, it is \emph{not} enhanced by the divergence of the superconducting density of states close to the gap.
For large detunings, the strong exchange field along the $z$ direction averages out the pseudospin accumulation along the $x$ and $y$ direction. As a consequence, the charge current tends to zero.

\begin{figure}
	\includegraphics[width=\columnwidth]{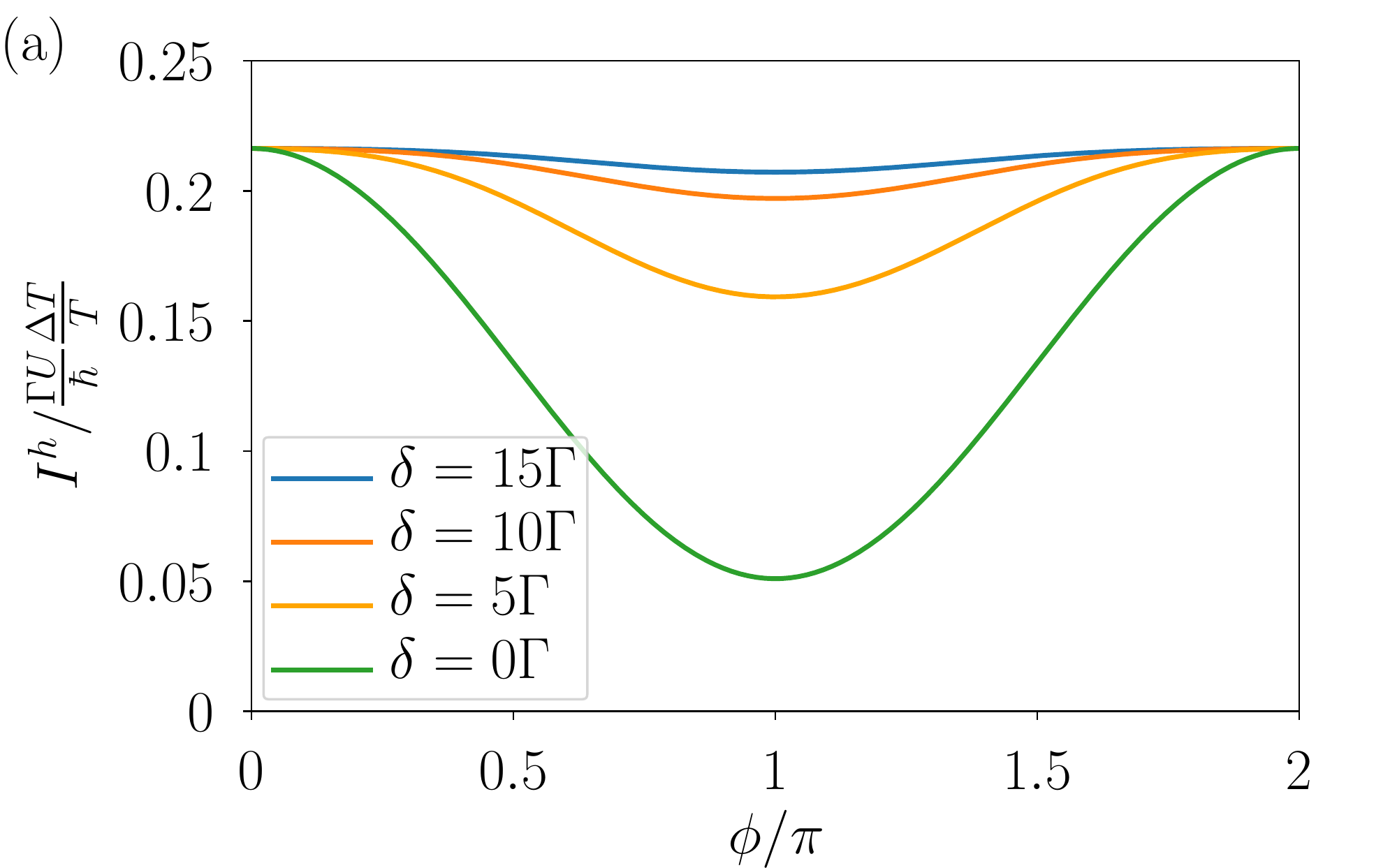}
	\includegraphics[width=\columnwidth]{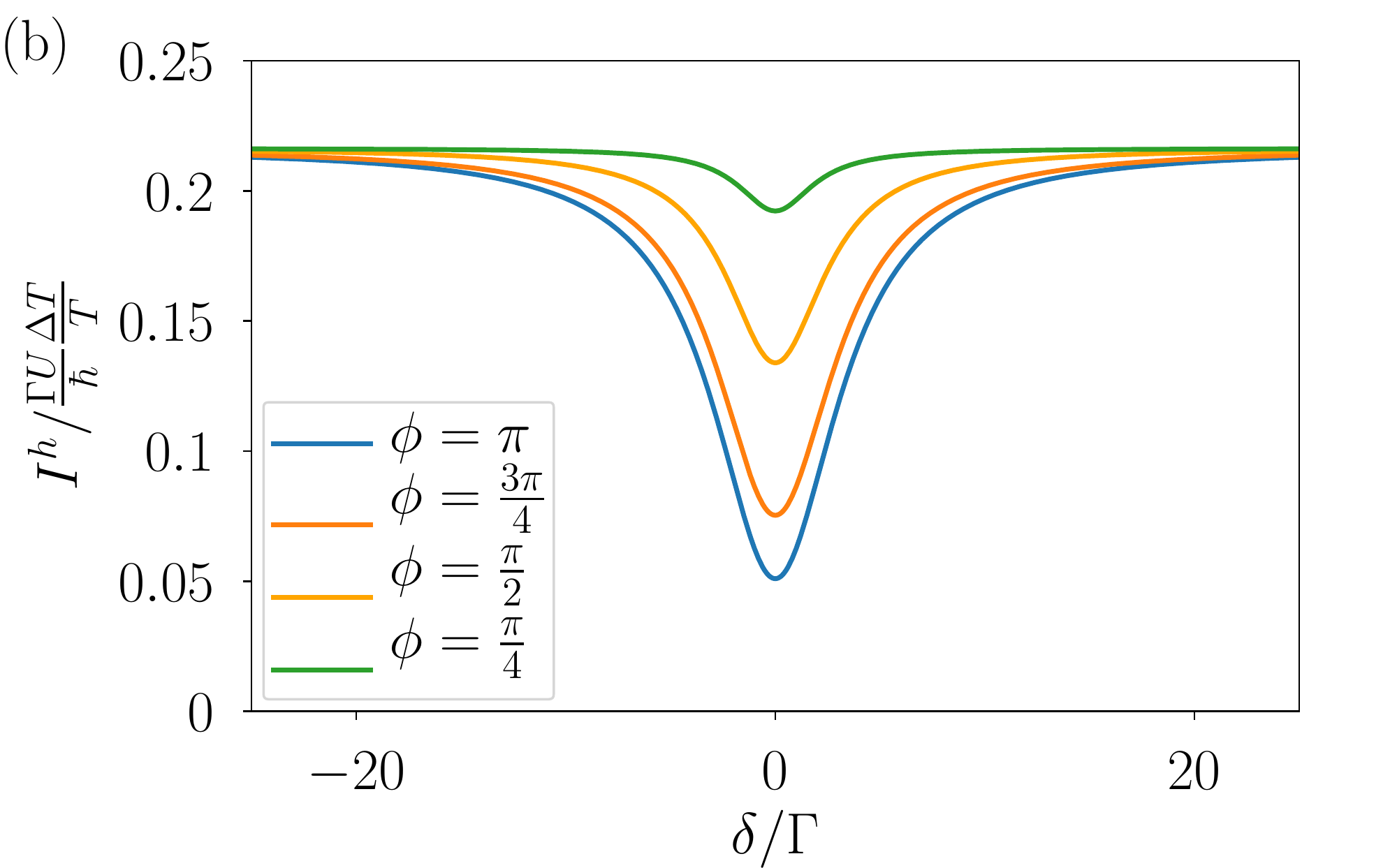}
	\caption{\label{fig:Ihline}Linear-response heat current $I^h$ as a function of (a) phase difference $\phi$ and (b) detuning $\delta$. Parameters as in Fig.~\ref{fig:Iclin}.}
\end{figure}
The heat current driven by a finite temperature bias $\Delta T$ is given by
\begin{equation}
	I^h=-\frac{U}{2}\left(Z^+_1+4I^{(1)}_yX_0^-\sin\frac{\phi}{2}\right)\frac{\Delta T}{T}.
\end{equation}
It consists of two contributions. The first one is independent of the phase difference $\phi$ and depends only on the tunnel coupling $\Gamma$, the Coulomb interaction $U$ and the superconducting order parameter $\Delta$. In contrast, the second contributions is sensitive to the phase difference $\phi$ and, thus, gives rise to a phase-coherent flow of heat which arises from the superconducting proximity effect on the dot. In consequence, it vanishes in the limit $\Delta\to0$.  Interestingly, the phase-dependent part of the heat current is proportional to $I^{(1)}_y$, i.e., it provides in principle direct information about the pseudospin accumulation on the dot. We remark that just like the charge current the heat current is also exponentially suppressed in $U/\kBT$. At the same time, however, it is enhanced by the increased superconducting density of states close to the gap. Hence, for the system heat currents in units of $\Gamma U/\hbar$ tend to be much larger than charge currents in units of $e\Gamma/\hbar$.

The phase dependence of the heat current is shown in Fig.~\ref{fig:Ihline}(a). At $\phi=0$, the heat current is maximal and takes the value $I^h=-UZ_1^+ \Delta T/(2T)$. The minimal heat current occurs at $\phi=\pi$ since $X^0_-$ is negative while the pseudospin accumulation $I^{(1)}_y$ is positive. This $\phi$ dependence of the thermal conductance differs from that of a tunneling Josephson junction which exhibits a maximum of the thermal conductance at $\phi=\pi$~\cite{maki_entropy_1965,maki_entropy_1966}. It rather resembles the phase-dependent thermal conductance of a transparent or topological Josephson junction which also has a minimum at $\phi=\pi$~\cite{zhao_phase_2003,zhao_heat_2004,sothmann_fingerprint_2016}. The ratio between the minimal and maximal heat current is given by $1-4\Delta^2/U^2$, i.e., it can be maximized by tuning the superconducting gap via the average temperature to be close to the Coulomb energy $U$. At the same time, this is also the regime where the relative modulation of the heat current becomes largest.

The $\delta$ dependence of the heat current is depicted in Fig.~\ref{fig:Ihline}(b). The largest modulation of the heat current occurs for $\delta=0$. In this case, the exchange field component along the $z$ axis vanishes which would otherwise reduce $I^{(1)}_y$ and thus the modulation amplitude. For the same reason, the modulation of the thermal conductance is strongly suppressed for large detunings $\delta\gg\Gamma$.

\subsection{\label{ssec:nonlinear}Nonlinear response}
\begin{figure}
	\includegraphics[width=\columnwidth]{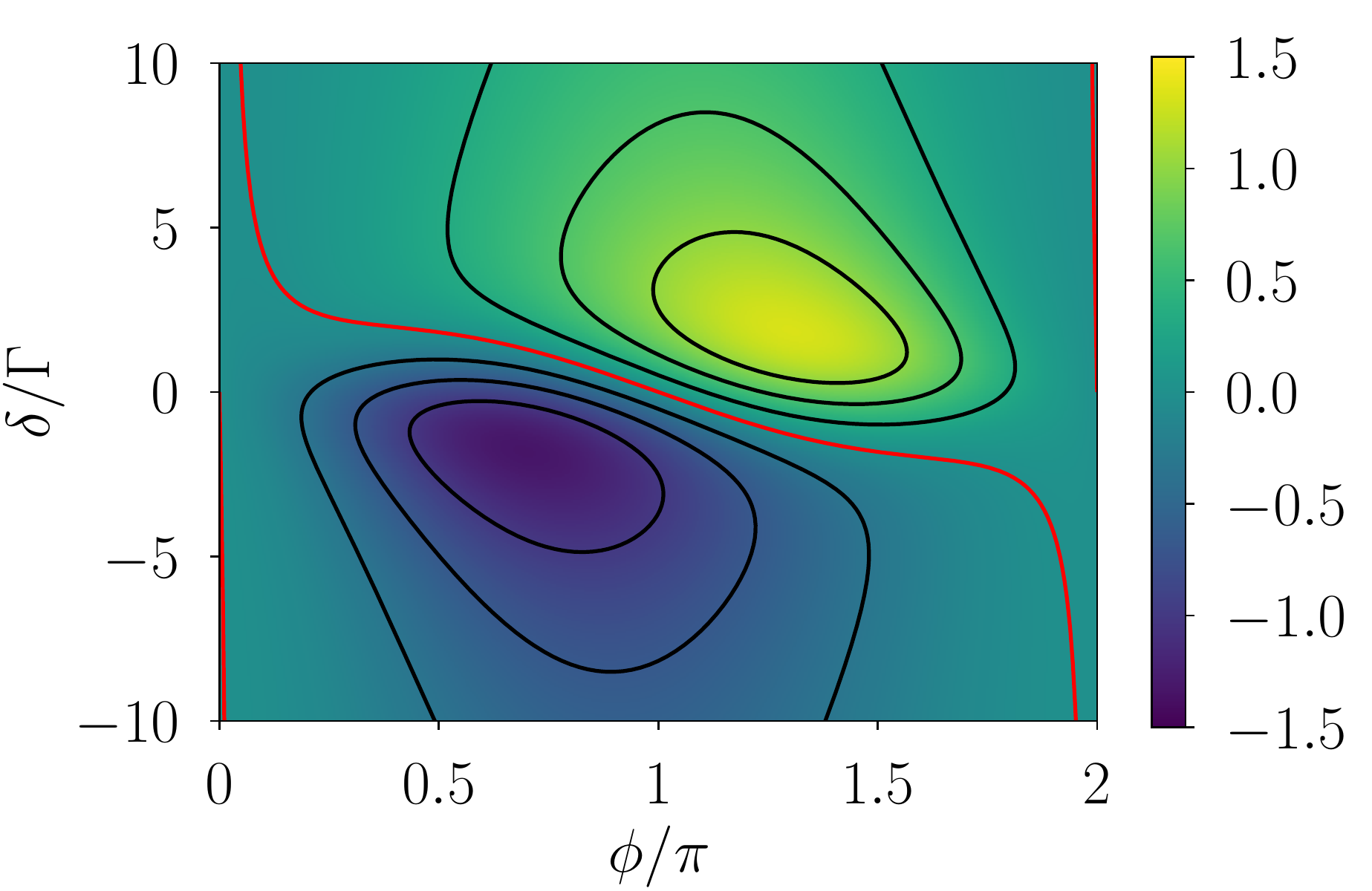}
	\caption{\label{fig:Icnonlinear}Charge current $I^e$ in units of $10^{-3}e\Gamma/\hbar$ as a function of phase difference $\phi$ and detuning $\delta$. The red line indicates a vanishing charge current. Parameters are $\Delta_0=2.32\kBT_\text{L}$, $U=5\kBT_\text{L}$, $\Gamma_\text{R}=4\Gamma_\text{L}$ and $T_\text{R}=T_\text{L}/2$.}
\end{figure}

We now turn to a discussion of transport in the nonlinear regime where a large temperature bias is applied across the system. The resulting charge current is shown as a function of phase difference and detuning in Fig.~\ref{fig:Icnonlinear}. Interestingly, for a phase difference $\phi\neq 0,\pi$, there is a finite charge current at the particle-hole symmetric point $\delta=0$.

This finite thermoelectric effect can be understood as follows. If the dot is empty (doubly occupied), electrons can virtually tunnel on (off) the dot and back. These virtual tunneling events give rise to a renormalization of the dot level energies which is captured by the real-time diagrammatic technique. Importantly, in the presence of Coulomb interactions, the renormalization is different for the empty and doubly occupied state and, thus, can break particle-hole symmetry effectively.
Hence, similarly to charge transport in quantum-dot spin valves~\cite{konig_interaction-driven_2003,braun_theory_2004,hell_spin_2015}, thermoelectric effects in superconductor-quantum dot hybrids constitute an important case where interaction-induced renormalization effects have a drastic impact on transport properties.
Using Eq.~\eqref{eq:Ic} the condition for a vanishing current can be cast into the compact form
\begin{equation}
	\frac{Z^-_\text{L}}{Z^-_\text{R}}=\frac{B_\text{L}\sin\left(\varphi-\frac{\phi}{2}\right)}{B_\text{R}\sin\left(\varphi+\frac{\phi}{2}\right)},
\end{equation}
where $\varphi$ denotes the $\delta$-dependent angle between the pseudospin and the $x$ axis. It illustrates the interplay between pseudospin relaxation and precession that influences the nonlinear charge current in a nontrivial way and is indicated by the red line in Fig.~\ref{fig:Icnonlinear}.

The nonlinear heat current behaves qualitatively similar to the linear-response case, i.e., it exhibits a minimum at phase difference $\phi=\pi$ and detuning $\delta=0$. We remark that the amplitude of the heat current oscillation is reduced in the nonlinear regime because the heat current at $\phi=\pi$ increases stronger with the temperature bias than the heat current at $\phi=0$.

\begin{figure}
    \includegraphics[width=\columnwidth]{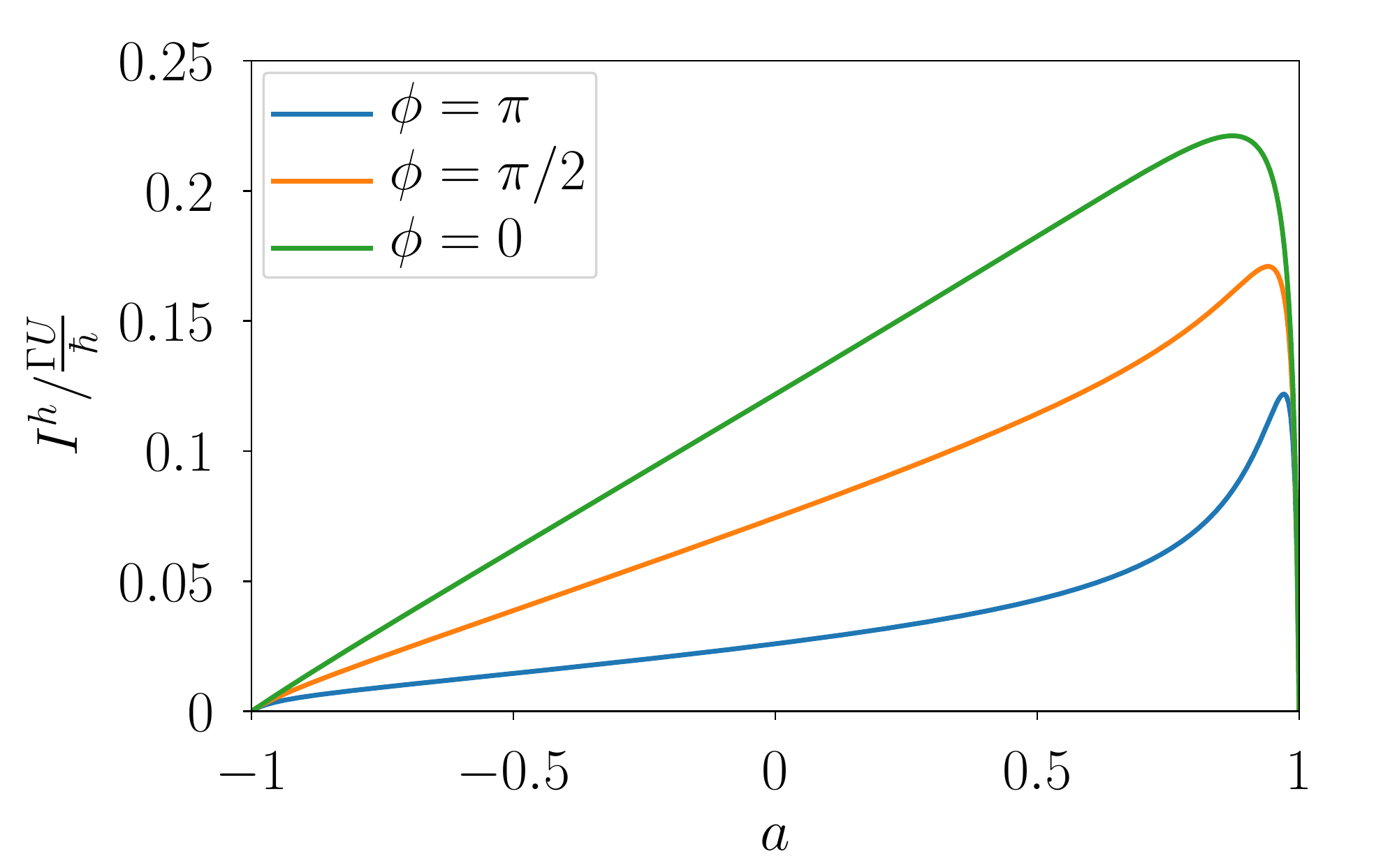}
	\caption{\label{fig:heatdiode}Nonlinear heat current as a function of the asymmetry $a$. Parameters are $\Delta_0=2.32 \kBT_\text{L}$, $U=4.64\kBT_\text{L}$, $\delta=10\Gamma$ and $T_\text{R}=0.1T_\text{L}$.}
\end{figure}

In the nonlinear regime, an asymmetric quantum-dot setup with $\Gamma_\text{L}\neq\Gamma_\text{R}$ can act as a thermal diode where the heat currents in the forward and backward direction are different. To discuss this effect in more detail, we introduce the asymmetry of tunnel couplings as $a=(\Gamma_\text{L}-\Gamma_\text{R})/(\Gamma_\text{L}+\Gamma_\text{R})$. The heat current in the forward direction is given by $I^h(a)$ while in the backward direction it is given by $I^h(-a)$. This definition is equivalent to denoting the forward (backward) direction as the one for which $T_\text{L}>T_\text{R}$ ($T_\text{L}<T_\text{R}$) at fixed tunnel couplings as long as $\Delta_{0,\text{L}}=\Delta_{0,\text{R}}$.

Figure~\ref{fig:heatdiode} shows the nonlinear heat current as a function of the asymmetry parameter $a$. For negative values of $a$, the heat current increases with $a$ while for positive values of $a$ it has a pronounced maximum. This nontrivial dependence on $a$ is most pronounced when the Coulomb energy is slightly larger than the superconducting gap. 
Since the heat current is not an even function of $a$, the system can rectify heat with a large heat current in the forward direction and a small heat current in the backward direction. For the chosen parameters we find that rectification efficiencies $I^h(a)/I^h(-a)\approx50$ can be achieved at the maximum forward heat current.

In order to understand the mechanism behind the thermal rectification, let us first consider the case of a single-level quantum dot coupled to two normal metal electrodes. At the particle-hole symmetric point, the heat current depends on the tunnel couplings via $\Gamma_\text{L}\Gamma_\text{R}/(\Gamma_\text{L}+\Gamma_\text{R})$. Hence, the heat current is an even function of the asymmetry $a$, $I^h(+a)=I^h(-a)$, such that thermal rectification does not occur.

For the superconducting system, the dependence of the heat current on the tunnel barriers is modified by the BCS density of states and is given by
\begin{equation}
	\frac{\Gamma_\text{L}\Gamma_\text{R}}{\Gamma_\text{L}\sqrt{U^2-4\Delta_\text{R}^2}+\Gamma_\text{R}\sqrt{U^2-4\Delta_\text{L}^2}}.
\end{equation}
Hence, due to the temperature dependence of the superconducting gap the heat current exhibits a nontrivial dependence on the asymmetry $a$ which forms the basis of the heat rectification mechanism. In addition, the coherent pseudospin dynamics of the dot can enhance the thermal diode effect for a finite phase difference $\phi$. As can be seen in Fig.~\ref{fig:heatdiode} it can increase the rectification efficiency by nearly a factor of 4 if the tunnel coupling asymmetry is adjusted to maximize the heat current in the forward direction. We remark that the enhancement of the rectification efficiency comes at the price of a slightly reduced heat current in the forward direction compared to the case $\phi=0$.

\section{\label{sec:conclusion}Conclusions}
We have analyzed thermally-driven transport through a superconductor-quantum dot hybrid in the sequential tunneling regime. We find that in linear response a finite thermoelectric effect can be generated close to the particle-hole symmetric point due to the superconducting proximity effect on the dot. In addition, there is a phase-dependent heat current through the quantum dot which in linear response is sensitive to the pseudospin accumulation in the dot, i.e., it provides direct access to information about the proximity effect on the dot.
In nonlinear response, an interaction-induced level renormalization due to virtual tunneling  gives rise to a finite thermoelectric response at the particle-hole symmetric point. Furthermore, the system can act as a thermal diode which is based on the temperature-dependence of the superconducting gap as well as the superconducting proximity effect.

Finally, we comment on potential experimental realizations of our proposal. For superconducting electrodes based on Al, the zero-temperature gap is given by \unit[0.17]{meV}, while the critical temperature is \unit[1.2]{K}. Hence, the device should be operated at temperatures around \unit[500]{mK} while the Coulomb interaction should be of the order of \unit[0.5]{meV}. Assuming furthermore tunnel couplings of the order of \unit[1]{Ghz}, we estimate charge currents of the order of \unit[0.1]{pA} and heat currents of the order of \unit[1]{pW} which are both within the reach of present experimental technology~\cite{fornieri_nanoscale_2016,timossi_phase-tunable_2018,dutta_thermal_2017}.

\acknowledgments
We thank Fred Hucht for valuable discussions and Stephan Weiss and Sun-Yong Hwang for feedback on the manuscript. We acknowledge financial support from the Ministry of Innovation NRW via the ``Programm zur Förderung der Rückkehr des hochqualifizierten Forschungsnachwuchses aus dem Ausland''.
This research was supported in part by the National Science Foundation under Grant No. NSF PHY-1748958.

\appendix
\section{\label{app:RTD}Real-time diagrammatics}
\begin{figure}
	\includegraphics[width=.2\textwidth]{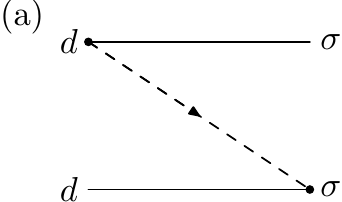}\hfill
	\includegraphics[width=.2\textwidth]{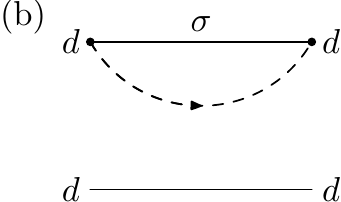}
	\includegraphics[width=.2\textwidth]{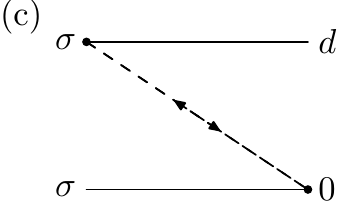}\hfill
	\includegraphics[width=.2\textwidth]{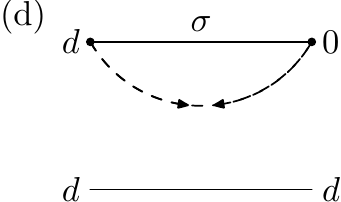}
	\caption{\label{fig:diagrams}Diagrams corresponding to different transitions in our setup system. Horizontal lines describe the forward and backward propagation of the dot on the Keldysh contour. Dots indicate tunneling vertices. Dashed lines correspond to tunneling lines which arise from Wick contractions of reservoir operators. Due to the presence of superconducting leads there are both normal (a), (b) and anomalous (c), (d) tunneling lines.}
\end{figure}

In this Appendix, we discuss the connection between real-time diagrams and the underlying physical processes. For a details on the diagrammatic theory for superconducting systems we refer the reader to Ref.~\cite{governale_erratum:_2008}.

Real-time diagrams consist of horizontal lines describing the forward and backward propagation of the quantum dot along the Keldysh contour. Dots on the Keldysh contour correspond to tunneling vertices where an electron is created (annihilated) on the dot and annihilated (created) in one of the superconductors. When we integrate out the noninteracting lead degrees of freedom, pairs of tunneling vertices get connected by tunneling lines. In superconducting systems, two different types of tunneling lines arise (i) normal lines which connect a vertex that creates an electron on the dot with a vertex that annihilates a dot electron and (ii) anomalous lines where two vertices that both annihilate (create) a dot electron are connected. The anomalous lines arise because the BCS Hamiltonian is diagonalized by Bogoliubov quasipartices which are superpositions of electrons and holes. Physically, they describe Andreev reflection processes where two electrons on the dot are created (annihilated) while a Cooper pair in the superconductor is annihilated (created).

Let us now focus on first order diagrams as depicted in Fig.~\ref{fig:diagrams}. Diagrams such as the one in Fig.~\ref{fig:diagrams}(a) describe the transition between two diagonal density matrix elements. They correspond to the usual transition rates that are obtained via Fermi's golden rule in conventional rate equation approaches. Diagrams such as shown in Fig.~\ref{fig:diagrams}(b) yield the diagonal elements of the rate matrix. While in rate equation approaches they are typically set by hand  to be $W_{\chi,\chi}^{\chi,\chi}=-\sum_{\chi'\neq\chi}W_{\chi',\chi}^{\chi',\chi}$ in order to ensure the conservation of probability, they appear naturally in the diagrammatic framework and, thus, provide an additional consistency check of the results.

In superconducting systems, additional diagrams involving anomalous tunneling lines such as the ones depicted in Fig.~\ref{fig:diagrams}(c) and (d) appear. They give rise to finite off-diagonal density matrix elements describing coherent superpositions of the dot being empty and doubly occupied and, hence, capture the superconducting proximity effect on the quantum dot. We emphasize that the proximity effect occurs already in first order in the tunnel coupling via these diagrams as they give rise to the coherent transfer of a Cooper pair between the dot and the superconductor. However, the proximity effect on the dot does not give rise to a supercurrent through the system in first order. A finite supercurrent relies on the coherent coupling between the two superconducting leads which for our setup can occur only in second- and higher-order processes. This is different in the case of a simple superconducting tunnel junctions where a finite supercurrent occurs already in first order~\cite{josephson_possible_1962}. Diagrams such as Fig.~\ref{fig:diagrams}(d) give rise to a level renormalization of the empty and doubly occupied state relative to each other and, thus, contribute to the exchange field in Eq.~\eqref{eq:Bex}.

\section{\label{app:Bex}Exchange field integral}
The integral appearing in the expression for the exchange field~\eqref{eq:Bex} can be solved analytically by performing the substitution $\omega\sign\omega=\Delta\cosh\alpha$. 
Subsequently, the residue theorem can be applied to the rectangle with corner points $(-R,R,R+2\pi i, -R+2\pi i)$ and taking the limit $R\to\infty$. While the contribution from the vertical edges vanishes, the top and bottom edge yield identical contributions. This allows us to express the exchange field integral as the infinite sum
\begin{widetext}
\begin{equation}
	B_\eta=\sum_{n=0}^\infty8\Gamma_\eta \kBT_\eta\frac{U}{[4(2n+1)^2\pi^2\kBT_\eta^2+U^2]}\frac{\Delta_\eta}{\sqrt{(2n+1)\pi^2\kBT_\eta^2+\Delta_\eta^2}}.
\end{equation}
\end{widetext}
For our numerical results, we have evaluated the sum by taking into account the first 10.000 summands.


%

\end{document}